\documentclass[a4paper,12pt]{article}
\usepackage{amsmath}    % need for subequations
\usepackage{graphicx}   % need for figures
\usepackage{verbatim}   % useful for program listings
\usepackage{color}      % use if color is used in text
\usepackage{float}
\usepackage{amsfonts}
\usepackage{algorithmic}
\usepackage{algorithm}
\usepackage{bm}
\usepackage{multirow}
\usepackage{amsthm}
\usepackage[latin1]{inputenc}
\usepackage{tikz}
\usetikzlibrary{shapes,arrows}
\usepackage[round]{natbib}
\usepackage{lscape}
\usepackage{rotating}
\usepackage{lineno}

\usepackage[final]{pdfpages}
\tikzstyle{decision} = [diamond, draw, fill=blue!20, 
    text width=4.5em, text badly centered, node distance=3cm, inner sep=0pt]
\tikzstyle{block} = [rectangle, draw, fill=blue!20, 
    text width=5em, text centered, rounded corners, minimum height=4em]
\tikzstyle{line} = [draw, -latex']
\tikzstyle{cloud} = [draw, circle, node distance=2.25cm,
    minimum height=3.5em]

\usepackage{authblk}

\newcommand{\latin}[1]{\textit{#1}}

\title{Quantifying uncertainty and dynamical changes in multi-species fishing mortality rates, catches and biomass by combining state-space and mechanistic multi-species models}
\author[1*]{Michael A. Spence}
\author[1]{Robert B. Thorpe}
\author[2]{Paul G. Blackwell}
\author[3]{Finlay Scott}
\author[4]{Richard Southwell}
\author[5]{Julia L. Blanchard}
\affil[1]{Centre for Environment, Fisheries and Aquaculture Science, Pakefield Road, Lowestoft, Suffolk NR33 0HT, UK}
\affil[2]{School of Mathematics and Statistics, University of Sheffield, UK}
\affil[3]{Oceanic Fisheries Programme, The Pacific Community (SPC), New Caledonia}
\affil[4]{Department of Mathematics, University of York, UK}
\affil[5]{Institute for Marine and Antarctic Studies and Centre for Marine Socioecology, University of Tasmania, Australia}
\affil[*]{michael.spence@cefas.co.uk}
\date{}
\begin{document}
%\linenumbers
\maketitle
%\textbf{Running title:} Avoiding the curse of circularity
\textbf{Running title:} Uncertainty in multi-species models
\newpage
\abstract{

In marine management, fish stocks are often managed on a stock-by-stock basis using single-species models. Many of these models are based upon statistical techniques and are good at assessing the current state and making short-term predictions; however, as they do not model interactions between stocks, they lack predictive power on longer timescales. Additionally, there are mechanistic multi-species models that represent key biological processes and consider interactions between stocks such as predation and competition for resources. Due to the complexity of these models, they are difficult to fit to data, and so many mechanistic multi-species models depend upon single-species models where they exist, or \latin{ad hoc} assumptions when they don't, for parameters such as annual fishing mortality.

In this paper we demonstrate that by taking a state-space approach, many of the uncertain parameters can be treated dynamically, allowing us to fit, with quantifiable uncertainty, mechanistic multi-species models directly to data. We demonstrate this by fitting uncertain parameters, including annual fishing mortality, of a size-based multi-species model of the Celtic Sea, for species with and without single-species stock-assessments. Consequently, errors in the single-species models no longer propagate through the multi-species model and underlying assumptions are more transparent. 

Building mechanistic multi-species models that are internally consistent, with quantifiable uncertainty, will improve their credibility and utility for management. This may lead to their uptake by being either used to corroborate single-species models; directly in the advice process to make predictions into the future; or used to provide a new way of managing data-limited stocks.}


\section{Introduction}

Food security has been highlighted as one of the major global challenges, with fisheries and aquaculture identified as key contributors to addressing this challenge \citep{FA02009,feeding_the_world}.
%As human populations has increased the demand on natural resources has increased. The marine environment has become a increasingly important resource 
%Food security and financial gains is important to the world ...
%exploitation 
%Therefore it is important to manage the marine environment.
Currently the majority of fish stocks are managed using single-species models (SSMs), such as the state-space assessment model (SAM) \citep{sam} and projections are made to assess the utility of management decisions.
Interacting stocks, which may compete with or predate on one another, can make conventional single-species management difficult \citep{TYRRELL20111,QUEROU2013192,Farcas_Rossberg}.
Alternatively a multi-species or whole ecosystem approach could be adopted to account for these interactions \citep{Pikitch346,link_2011,Plag_2014}. 
There are several multi-species models (MSMs) ranging from statistical models \citep[e.g.\ Stochastic MSM (SMS)][]{sms}, to more mechanistic-based models \citep[e.g.\ mizer;][]{mizer} or whole ecosystem models \citep[e.g.\ StrathE2E;][]{heath}. 
%Statistical models are good at accurately determining the current and historical states of stocks, whereas mechanistic-based models are better at forecasting the future states of the stocks \citep{Farcas_Rossberg,connor}.

%% --new bit 
SSMs and statistical MSMs are often used to describe the current and recent status of the system, and to make short-term forecasts. 
They aim to learn about the system by fitting many `tuning parameters', parameters that are adjusted to make the model look like the observed system \citep{Plag_2014,Brynjarsd_ttir_2014}.
On the other hand, mechanistic models, sometimes called process-based models, are based on the theoretical understanding of the relevant ecological processes \citep{Cuddington13}. They generally model the behaviour of the system through differential equations and/or a series of rules or algorithms.
They prioritise realism over reality, often explaining why things happen rather than describing what happened \citep{white_marshall}. %, which allows one to explore scenarios that are different to things that have previously happened.
Many of the parameters are treated as `input variables', with values taken from other sources \citep{Brynjarsd_ttir_2014}, leaving fewer `tuning parameters' that represent processes that are either too complex or not known, e.g. recruitment.
For example, in mechanistic size-based MSMs, the predator-prey mass ratio is an `input variable', coming from other studies \citep[e.g.][]{Hattonaac6284}, whereas in statistical MSMs it is treated as a `tuning parameter' and learned from data \citep[e.g.][]{wgsam2017} (see Supplementary material S5 for an illustrative example of `tuning parameters' and `input variables'). 
As mechanisms and physical laws are time invariant and more robust than statistical correlations, mechanistic models are better at predicting outside the immediate domain in which they were fitted, such as in the future \citep{connor,Cuddington13}. 

%They have been used as management tools in other areas such as  weather forecasting \citep{LYNCH20083431} and climate science \citep{IPCC}.

%However, their realism comes at an expense of accuracy. Some processes that are poorly understood or to complex to include in the mechanistic models, such as recruitment, and are replaced with `tuning` parameters.

%% end new bit

%% in discussion --there's it's okay to use tuning parameters as inputs but that has to be acknowledged

%%he advantage of using mechanistic models is that unrelated studies and data can be easily incorporated into the model, adding to the realism of the model.

Often mechanistic MSMs are fitted to, or rely on inputs from, SSMs \citep[e.g.][]{blanchard, mackinson}.
A common example is instantaneous fishing mortality values that are taken from SSMs, to drive fishing dynamics in MSMs \citep[e.g.][]{spence_ns,fishsums}.
In some ecoregions, fishing mortality values from SSMs either do not exist for all species or only qualitative patterns are reported. In studies with MSMs, fishing dynamics for species without fishing mortality values from SSMs are added using \latin{ad hoc} methods \citep{thorpe15}.
Further, as models are simplifications of reality and often the fishing mortality is treated as a `tuning parameter', the fishing mortality values lose their interpretation outside of the fitted model \citep{rougier2013}. Thus they are not the same as the true instantaneous fishing mortality values but instead are model specific.
%Fishing mortality rates are emergent properties of the system and depend not only on the level of the stock in question but also on the level of other stocks. These interactions are implicit in SSMs but are explicit in MSMs. Thus, the fishing mortality rates calculated from SSMs are not the same as the ones that should go into MSMs. 
For example statistical MSMs, that are often used to generate natural mortality values for SSMs, have different fishing mortality than the SSMs \citep[e.g. North Sea Cod in SMS and SAM;][]{wgsam2017,cod2018advice}, despite being fitted to the same data and having a similar representation of the population structure.
Fitting MSMs to SSMs or taking inputs from them can lead to circularity in results as errors propagate through the models \citep{brooks}.

In MSMs, fitting fishing can be a challenging task.
Recent software advances (e.g.\ ADMB \citep{admb}) have meant that statistical MSMs, designed with tractability in mind, are relatively easy to fit.
For mechanistic-based MSMs, %which generally model the behaviour of the system through a series of rules or algorithms, rather than statistical models, 
fitting using traditional methods can be a difficult task. Evaluating the output of a model for a particular set of inputs can often be done only by running the model, which can take anything from a few seconds to a few hours. This means that fitting a large number of uncertain parameters, such as fitting fishing mortality for each year, can be a difficult task.
Furthermore, for these models to be any use to support management, outputs need to be reported with robust estimates of uncertainty \citep{harwood_stokes}. 

Parameter uncertainty has previously been explored in MSMs to explore a handful of parameters \citep{thorpe15,mackinson}.
\citet{spence_ns} fitted a size-based model of the North Sea using a Bayesian framework, which we adopt here \citep{bayes}, using Markov chain Monte Carlo (MCMC) to sample from the posterior distribution \citep{metropolis,hastings}.
%For all but the simplest MSMs, the posterior distribution is intractable and so samples from it are often generated using Markov chain Monte Carlo (MCMC) \citep{metropolis}.
%can only be calculated up to a normalising constant. It is possible to set up a Markov chain, where new parameter sets are proposed and then accepted or rejected with some probability, that has a stationary distribution equal to the posterior distribution. This enables samples from the posterior distribution to be taken and is known as Markov chain Monte Carlo (MCMC) \citep{metropolis}.
%Therefore, fitting these models with many uncertain parameters is often not feasible.
Adding dynamical parameters, such as annual fishing mortality, makes the uncertain parameter space very large, making it difficult to explore. However, we may be able to consider the model as a state-space model, a common approach in SSMs \citep[see][for a recent review]{rev_ssm}. %, we can utilise the dependence structure to efficiently explore the dynamical parameters space. 
In state-space models, the `state' of the system is updated using a Markov process, known as the process model, and there are some noisy, possibly incomplete, observations of the `state', %known as the
defined by an observation model. State-space models have a specific dependence structure (see Figure \ref{fig:dag}), with the observations of the past and present being conditionally independent given the unobserved state, a structure that can be advantageous when fitting the model \citep{Zucchini}.

There are many methods of fitting non-linear state-space models including Extended Kalman Filers \citep{evensen,wan00}, MCMC methods \citep{jonsen05} and using the Laplace approximation \citep{tierney_kadane_86} to integrate out the unobserved states \citep{skaug_fournier_06}.
\citet{spence_ff} used particle filters \citep{gordon,liu_chen_98} to update a few years of fishing rates in two MSMs, but for longer periods of time this method is not practical. This is due to the likelihood being largely dominated by the process model and not the observation model which leads to poor mixing of the MCMC \citep{fasiolo2016}. In this paper we develop an MCMC algorithm that sequentially updates each dynamical parameter and improves the mixing of the MCMC.

In many cases the only way of evaluating the likelihood of parameter values is to run the model.
Running mechanistic models can be slow so ideally one would want to parallelise the model when fitting to data; however this is difficult for MCMC, as iterations need to be done sequentially \citep{parallel_mcmc}. 
Some MCMC algorithms have been developed that take advantage of parallel computing \citep{nicholls,caulderhead},
whereas others reduce the number of times that the model needs to be run. The delayed-acceptance MCMC algorithm \citep{sherlock} uses a fast approximation of the likelihood, possibly using a statistical model, before deciding whether or not to run the mechanistic model. Due to the high dimensionality of this problem, fitting accurate fast approximations of the likelihood can be difficult, but for many of these problems there are some parameters that affect only part of the likelihood. Here we introduce a second new MCMC algorithm that runs several proposals in parallel using the mechanistic model and then combines them to give a single proposal that has an increased chance of being accepted.

In this paper we fit fishing mortality and other uncertain parameters of a multi-species size-based model for the Celtic Sea, without the use of SSMs.  We compare stock-assessments made using the MSM with those developed using SSMs. 
Although demonstrated on a multi-species marine model, this problem is not unique to MSMs and methods demonstrated here can be used for fitting mechanistic-based models of intermediate complexity, e.g.\ individual-based models \citep{railsback_grimm}, especially when there are dynamic parameters.
In Section \ref{sec:methods} we define state-space models, describe the multi-species size-based model, the data and the fitting procedure as well as the two new MCMC algorithms. In Section \ref{sec:results} we describe the results of the fitted model and we conclude with a discussion in Section \ref{sec:discussion}. We also demonstrate the fitting procedure with a simulation study using another mechanistic MSM \citep{LeMaRns} in the Simulation study.

\section{Methods}
\label{sec:methods}
In this section we describe how we can treat the MSM as a state-space model. We introduce the MSM used in this study, the uncertain parameters, which include fishing mortality for each species for each year, and the data to which the model was fitted. We then describe the steps used to sample from the posterior distribution using Markov Chain Monte Carlo (MCMC).

\subsection{State-space model}

\begin{figure}[ht]
\begin{center}
\includegraphics{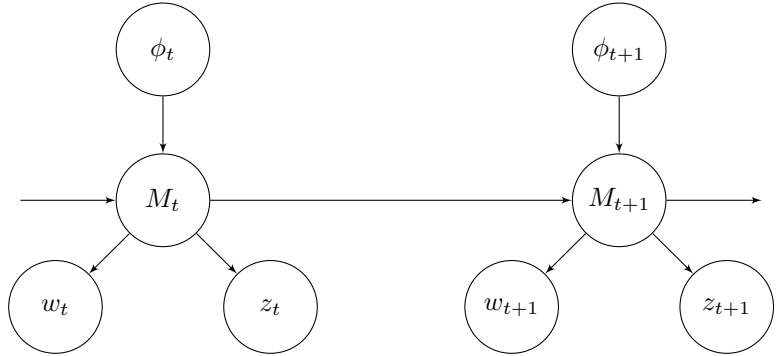}
\end{center}
\caption{A directed acyclic graph of the state-space model.}
\label{fig:dag}
\end{figure}

Let ${M}_t$ be the state of the MSM at time $t$. Then 
\begin{linenomath}
\begin{equation*}
{M}_{t}|{M}_{t-1} \sim h({M}_{t-1},\bm{\phi}_t,\bm{\theta}),
\end{equation*}
\end{linenomath}
where $\bm{\phi}_t$ are dynamical parameters at time $t$ and $\bm\theta$ are static parameters. $h(\cdot)$ is known as the process model. We do not observe the state directly but at time $t$ we observe $\bm{y}_t$, where
\begin{linenomath}
\begin{equation*}
\bm{y}_t|{M}_t\sim{}g({M}_t,\bm\sigma^2),
\end{equation*}
\end{linenomath}
and $\bm{\sigma}^2$ are static parameters. $g(\cdot)$ is known as the observation model. Figure \ref{fig:dag}
represents this model as a directed acyclic graph (DAG).

\subsubsection*{Process model}
The process model $h(\cdot)$ used here is the deterministic multi-species size-based model, mizer \citep{hartvig,mizer}, and the state of the model, $M_t$, is a collection of the density of all species, $N_i(m)$, and the background resource, $N_R(m)$ at all weights, $m$ at time $t$ (see Supplementary material S1 for details).
Mizer was developed to represent the size and abundance of all organisms from zooplankton to large fish predators in a size-structured food web. Some species are represented by species-specific traits and body size while others are represented solely by body size. The core of the model involves ontogenetic feeding and growth, mortality, and reproduction driven by size-dependent predation and maturation processes. The smallest individuals in the model do not eat fish belonging to the fish populations, but consume smaller planktonic or benthic organisms which we describe as a background resource spectrum. Fish grow and die according to size-dependent predation and, if mature, recruit new young which are put back into the system at the minimum weight. As well as the predation and background mortality, the fish in the model also experience fishing mortality. 

In this study we fit mizer for 17 species, shown in Table \ref{tb:spec}, in the Celtic Sea, ICES (International Council for Exploration of the Seas) areas 7e-j.
A description of the model can be found in the Supplementary material (S1) along with the parameter values.

%% add description of the fishing rates
In mizer there are a number of uncertain parameters to estimate. The carrying capacity of the background resource spectrum, $\kappa$, is uncertain, with a relatively uninformative prior distribution given by $\ln(\kappa)\in[0,40]$ uniformly (see Table \ref{tb:par}). Recruitment follows a density-dependent process with the maximum number of recruits of the $i$th species being $R_{max,i}$, which is also uncertain. We specified a relatively uninformative prior distribution as $\ln(R_{max,i})\in[0,50]$ uniformly (see Table \ref{tb:par}), for all $i$. The fishing mortality of the $i$th species of weight $m$ at time $t$ was
\begin{linenomath}
\begin{equation*}
\phi_{t,i}q_i(m),
\end{equation*}
\end{linenomath}
where $q_i(m)$ is the catchability of species $i$ at size $m$, normalised so that $\max_m(q_i(m))=1$, and $\phi_{t,i}$ is the fishing rate (values for $q_i(m)$ are shown in the Supplementary material (Figure S1)). The model was run from 1991-2014 ($t=1,\ldots,24$) and the fishing rate for each species for each year was also uncertain with $\phi_{t,i}\in[0,1.5]$ uniformly for $t=1,\ldots,24$ and for all $i$.

The model can be sensitive to its initial state, when $t=0$, and so the model was projected for 300 years to a stationary state, a process known as spin-up, with a fixed fishing rate $\phi_{0,i}$ for each species prior to running for $t=1,\ldots,24$. As in \citet{spence_ns} we treated the spin-up fishing rates as additional parameters with $\phi_{0,i}\in[0,1.5]$ uniformly for all $i$ (see Table \ref{tb:par}). We consider $\bm{\theta}=(\ln\kappa,\ln{}R_{max,1:17},\phi_{0,1:17})'$ to be `static' parameters and the fishing rates, $\phi_{1:24,1:17}$ to be `dynamical' parameters (with $1{:}17$ meaning $i=1\ldots17$). %% could work on this sentence and maybe the ordering

In addition to the commercial fishing mortality, we included survey fishing mortality. The catchability of the survey vessel was taken from \citet{walker_selection} and the fishing effort for the survey effort taken from DATRAS \citep{datras}. By including the survey fishing mortality we are able to fit the model to data from survey.

\begin{table}[H]
\caption{The species in the Celtic Sea mizer model}
\label{tb:spec}
\begin{tabular}{lll}
\hline
$i$ & Common name & Latin name  \\
\hline
1 & Atlantic herring & Clupea harengus\\
2 & European sprat & Sprattus sprattus\\
3 &Atlantic cod & Gadus morhua \\
4 & Haddock & Melanogrammus aeglefinus \\
5 & Whiting & Merlangius merlangus\\
6 & Blue whiting & Micromesistius poutassou\\
7 & Norway pout & Trisopterus esmarkii\\
8 & Poor cod &Trisopterus minutus \\
9 & European hake & Merluccius merluccius\\
10 & Monkfish & Lophius piscatorius\\ % check that this is the correct species
11 & Atlantic horse mackerel & Trachurus trachurus\\
12 & Atlantic mackerel & Scomber scombrus\\
13 & Common dab & Limanda limanda\\
14 & European Plaice & Pleuronectes platessa\\
15 & Megrim & Lepidorhombus whiffiagonis\\
16 & Common sole & Solea solea\\
17 & Boarfish & Capros aper
\end{tabular}
\end{table}

\begin{sidewaystable} \centering 
\caption{The uncertain parameters.}
\label{tb:par}
\begin{tabular}{lllll}
\hline
Parameters & Dimensions & Units & Prior & Notes\\
\hline
$\ln{R_{max,1:17}}$ & 17 & $\ln(vol^{-1}grams^{-1}year^{-1})$  &$U(0,50)$& Natural log of the maximum recruitment\\
 &&&& for each species\\
$\ln{\kappa}$ & 1 & $\ln(grams^{-\lambda-1}vol^{-1})$ & $U(0,40)$ &Natural log of the carrying capacity \\
 &&&& of the resource spectrum\\
$\phi_{0,1:17}$ & 17 & $year^{-1}$ & $U(0,1.5)$& The fishing rates during the spin-up\\
 &&&& period for each species\\
$\phi_{1:24,1:17}$ & $17\times24=408$ & $year^{-1}$& $U(0,1.5)$&The fishing rate for each species\\
  &&&& for each year\\
$\sigma^2_{s,1:17}$ & 17 & Unitless & $Inv-Gamma(2,2)$& The variance of the error on \\ 
 &&&& the natural log survey catches\\
$\sigma^2_{c,1:17}$ & 17 & Unitless & $Inv-Gamma(0.1,0.1)$& The variance of the error on\\ 
 &&&& the natural log commercial catches\\
\end{tabular}
\end{sidewaystable}

\subsubsection*{Observation model}
At time $t$, we observe catches in tonnes, $\bm{y}$, made up of those by commercial vessels, $\bm{w}_t$ for $t=1,\ldots{},24$ (1991-2014), and those by the International Bottom Trawl Survey (IBTS), $\bm{z}_t$ for $t=7,\ldots{},24$ (1997-2014), with $|\bm{w}_t|=|\bm{z}_t|=17$. We take
\begin{linenomath}
\begin{equation*}
\ln\bm{w}_t\sim{}N(\ln{\bm{c}(M_t)},\Sigma_c)
\end{equation*}
\end{linenomath}
where $\bm{c}(M_t)$ is the commercial catch from the process model and $\Sigma_c$ is a diagonal matrix with elements $\bm{\sigma}_c^2$. Similarly we take 
\begin{linenomath}
\begin{equation*}
\ln\bm{z}_t\sim{}N(\ln{\bm{s}(M_t)},\Sigma_s)
\end{equation*}
\end{linenomath}
where $\bm{s}(M_t)$ is the survey catch from the process model and $\Sigma_s$ is a diagonal matrix with elements $\bm{\sigma}_s^2$. The $i$th elements of $\bm{c}(M_t)$ and $\bm{s}(M_t)$ are denoted ${c}(M_t)_i$ and ${s}(M_t)_i$ and defined in equations S3 and S4 in the Supplementary material respectively.
The likelihood of the model is
\begin{linenomath}
\begin{eqnarray}
l(\bm{y}|\bm{\theta},\phi_{1:24,1:17},\bm{\sigma}_c^2,\bm{\sigma}_s^2)&=&
\prod_{i=1}^{17}\prod_{t=1}^{24}N(\ln(w_{t,i})|\ln(c(M_t)_i),\sigma_{c,i}^2)\nonumber\\
&&\quad\times\prod_{t=7}^{24}N(\ln(z_{t,i})|\ln(s(M_t)_i), \sigma_{s,i}^2),
\label{eq:full_like}
\end{eqnarray}
\end{linenomath}
where $w_{t,i}$, $z_{t,i}$, $\sigma_{c,i}^2$ and $\sigma_{s,i}^2$ are the $i$th element of $\bm{w}_t$, $\bm{z}_t$, $\bm{\sigma}_c^2$ and $\bm{\sigma}_s^2$ respectively, and $N(a|d,e)$ is a normal density with expectation $d$ and variance $e$ evaluated at $a$.
Table \ref{tb:par} summarises the uncertain parameters.

\subsection{Data}
Landings data were extracted from ICES \citep{ices_landings} and discards were estimated as a percentage of the retained biomass \citep{HEYMANS2016173,discards}. All discards were assumed to have been removed from the living stock in the process model, such that all discards are assumed to have died. As only discards and no landings were recorded for poor cod and Norway pout, we fixed the variance of the commercial catches, $\sigma^2_{c,7:8}=(4,4)'$ \citep{shepard}.
We extracted the IBTS survey data from DATRAS \citep{datras} from 1997 until 2014 (t=7,\ldots,24).

\subsection{Fitting the model}
The model was fitted in a Bayesian framework so that we could quantify the uncertainty in the model parameters using probability. As the likelihood was intractable we were required to sample from the posterior distribution. Although a suitable Markov Chain with stationary distribution equal to the posterior would eventually converge to the posterior distribution, this would take a long time. To speed the process up we aimed to start the Markov chain close to the high-probability region of the posterior distribution. To find these starting values we used history matching to reduce the parameter space \citep{veron}.

%\subsubsection*{History matching}
%History matching is a method of excluding implausible parameter space, space that has negligible posterior density, in order to find the area of parameter space with non-negligible posterior density. Using Sobol sequences \citep{sobol}, a space filling algorithm, to sample possible parameter and fishing mortality values, we were able to run the model 5000 times.
%The modelled catches were fitted to the parameter values using generalised additive models (GAMs) \citep{wood_gam}. These models allow the catches of the full model to be predicted, with uncertainty, for all possible values of the parameters. Using the GAMs, we excluded regions of space that predicted mean catches that were significantly different from the observed mean catches.
%This process was repeated until we found a parameter set that was sufficiently close to the observation.

\subsubsection*{Markov Chain Monte Carlo}
The posterior distribution was explored using MCMC. 
Due to the high dimensionality of the parameter space, mixing efficiently was difficult and so we developed two extensions of the delayed-acceptance MCMC algorithm of \citet{sherlock} that take advantage of parallel computing and explore the posterior distribution in an efficient way.

The first extension, which we refer to as the marginal-delayed-acceptance MCMC (MDA-MCMC), is shown in Algorithm \ref{alg:mda}.
\begin{algorithm}
\caption{An iteration of the marginal-delayed-acceptance MCMC algorithm (MDA-MCMC).
The current parameters $\bm\theta$, are divided into $N+1$ disjoint sets with the $i$th set being denoted $\bm{\theta}_i$, having the likelihood evaluation $l_i(\bm{y}|\bm\theta)$ and proposal distribution $f_i(\cdot|\bm{\theta}_i)$. $p(\bm{\theta})$ is the prior and $l(\bm{y}|\bm\theta)$ is the full likelihood. We define $\land$ to be the minimum, i.e. $a\land{}b=\min(a,b)$.}
\label{alg:mda}
\begin{algorithmic}
\STATE $\bm{\theta}''\gets\bm{\theta}$
\FOR{$i$ in $1:N$}
\STATE $\bm{\theta_i}' \sim{} f_i(\cdot|\bm\theta_i)$
\STATE $\bm{\theta_i}'' \gets\bm{\theta_i}' $ with probability
\begin{equation*}
\alpha_i(\bm\theta,\bm\theta'_i)=1\land
\frac{p(\bm\theta'_i)l_i(\bm{y}|{\bm\theta}'_i,\bm{\theta}_{-i})}{p(\bm\theta_i)l_i(\bm{y}|\bm{\theta})}
\end{equation*}
\ENDFOR
\STATE $\bm{\theta} \gets \bm{\theta}''$ with probability
\begin{eqnarray*}
1\ \land
&&\frac{p(\bm\theta'')l(\bm{y}|\bm{\theta}'')}{p(\bm\theta)l(\bm{y}|\bm{\theta})}\\
&\times&\prod_{\{i:\bm\theta''_i\neq\bm\theta_i\}}\frac{f_i(\bm\theta_i|\bm\theta'_i)}{f_i(\bm\theta'_i|\bm\theta_i)}\times\frac{\alpha_i(\bm\theta'',\bm\theta_i)}{\alpha_i(\bm\theta,\bm\theta_i')}\\
&\times&\prod_{\{i:\bm\theta''_i=\bm\theta_i\}}\frac{1-\alpha_i(\bm\theta'',\bm\theta_i')}{1-\alpha_i(\bm\theta,\bm\theta_i')}
%\right\}
\end{eqnarray*}
\end{algorithmic}
\end{algorithm}
It is understood that when moving in lower dimensions it is possible to make larger moves \citep{neal2006}; here we propose several moves in smaller dimensions and check their suitability before trying to make the move itself. 
For each iteration the parameter set is divided into $N+1$ disjoint sets with $N$ of the sets each having some likelihood function, $l_i(\cdot)$, associated with it. This algorithm attempts to update the parameters in the first $N$ sets whilst holding the parameters in the $N+1$ set, which may be empty, fixed.
$N$ of the parameter sets are each updated by one iteration of the Metropolis-Hastings MCMC algorithm, keeping the other parameters fixed, with its own likelihood function. If the current model run is saved, this would cost $N$ new model evaluations ($N+1$ if not) that could be done in parallel and so could, in terms of clock time, take one model evaluation. The output from each of the $N$ MCMC algorithms is used as a proposal for the main MCMC algorithm. This then takes a further two new model evaluations which could be performed in parallel. Using the acceptance rates described in Algorithm \ref{alg:mda} leads to a Markov Chain with the correct stationary distribution, a proof of which is in the Supplementary material (S3).

The second extension, which we call particle-delayed-acceptance MCMC (PDA-MCMC), is shown in Algorithm \ref{alg:pmda}. 
\begin{algorithm}
\caption{An iteration of the particle-delayed-acceptance MCMC algorithm (PDA-MCMC). Let $M_t=h(M_{t-1},{\phi}_{1:17,t},\bm{\theta})$ be the model run up until time $t$, with $M_0$ being its initial state and $k_t(M_t)$ be a likelihood evaluation of this model. The static parameters are $\bm{\theta}$, the current fishing rates are ${\phi}_{1:17,1:24}$ and $f(\cdot|\phi_{1:17,t})$ is the proposal distribution. The full likelihood is $l(\bm{y}|{\phi}_{1:17,1:24})$ and $p(\phi_{1:17,1:24})$ is the prior. We define $\land$ to be the minimum, i.e. $a\land{}b=\min(a,b)$.}
\label{alg:pmda}
\begin{algorithmic}
%\COMMENT Run the model up until $M_0$.
\STATE $Q_0\gets{}M_0$, $\phi''_{1:17,1:24}\gets{}\phi_{1:17,1:24}$
\FOR {$t$ in $1:24$}
\STATE ${\phi}_{1:17,t}'\sim{}f(\cdot|{\phi}_{1:17,t})$
\STATE $M'_t\gets{}h(M_{t-1},{\phi}_{1:17,t}',\bm{\theta})$ and 
$M_t\gets{}h(M_{t-1},{\phi}_{1:17,t},\bm{\theta})$ 
\STATE $Q'_t\gets{}h(Q_{t-1},{\phi}_{1:17,t}',\bm{\theta})$ and $Q_t\gets{}h(Q_{t-1},{\phi}_{1:17,t},\bm{\theta})$
\STATE \item ${\phi}_{1:17,t}'' \gets{\phi}_{1:17,t}'$ and $M_t\gets{}M_t'$ with probability
\begin{equation*}
\alpha_t({\phi}_{1:17,t},{\phi}_{1:17,t}')=1\land
\frac{p(\phi_{1:17,t}')k_t(M'_t)}{p(\phi_{1:17,t})k_t(M_t)}
\end{equation*}
\ENDFOR
\STATE
${\phi}_{1:17,1:24}\gets{\phi}''_{1:17,1:24}$ with probability
\begin{eqnarray*}
1\ \land&&
\frac{p(\phi_{1:17,1:24}'')l(\bm{y}|{\phi}_{1:17,1:24}'')}{p(\phi_{1:17,1:24})l(\bm{y}|{\phi}_{1:17,1:24})}\\
&\times&
\prod_{\{t:\bm{}\phi_{1:17,t}\neq\bm{}\phi''_{1:17,t}\}}\frac{f(\phi_{1:17,t}|\phi_{1:17,t}')}{f(\phi_{1:17,t}'|\phi_{1:17,t})}\times\frac{
1\land
\frac{p(\phi_{1:17,t})k_t(Q_t)}{p(\phi_{1:17,t}')k_t(Q_t')}
}{\alpha({\phi}_{1:17,t},{\phi}_{1:17,t}')}\\
&\times&
\prod_{\{t:\phi_{1:17,t}=\phi''_{1:17,t}\}}\frac{1 - 
1\land
\frac{p(\phi_{1:17,t}')k_t(Q_t')}{p(\phi_{1:17,t})k_t(Q_t)}
}{1 - \alpha({\phi}_{1:17,t},{\phi}_{1:17,t}')}
\end{eqnarray*}
\end{algorithmic}
\end{algorithm}
In PDA-MCMC the fishing rates for each year are sequentially updated using the Metropolis-Hastings algorithm. 
Once the algorithm has updated for each year of the model, the new fishing rates are used as a proposal for the MCMC update. This requires five model runs, which could be as quick as two model runs in terms of clock time (as the four of the model runs could be parallelised) and leads to a Markov Chain with the correct stationary distribution, a proof of which is in the Supplementary material (S3).

To sample from the whole posterior distribution we used a random walk Metropolis-within-Gibbs algorithm with proposal variances tuned from a pilot run. At each iteration we performed four types of updates:
\begin{enumerate}
\item Update $\ln{R_{max,1:17}}$ and $\phi_{0,1:17}$ together using the MDA-MCMC algorithm with $N=17$.
The $i$th set was $\{\ln{R_{max,i}},\phi_{0,i}\}$ with 
\begin{linenomath}
\begin{equation*}
    l_i(\bm{y}|\bm\theta)=\prod_{t=1}^{24}N(\ln(w_{t,i})|\ln(c(M_t)_i),\sigma_{c,i}^2)\\
\times\prod_{t=7}^{24}N(\ln(z_{t,i})|\ln(s(M_t)_i), \sigma_{s,i}^2)
\end{equation*}
\end{linenomath}
and the full likelihood, $l(\bm{y}|\bm{\theta})$ being $l(\bm{y}|\bm{\theta},\phi_{1:24,1:17},\bm{\sigma}_c^2,\bm{\sigma}_s^2)$ from equation \ref{eq:full_like}. The 18th set, which does not get updated at this step, was $\{\ln{(\kappa),\phi_{1:17,1:24},\bm{\sigma}_{c}^2,\bm{\sigma}_{s}^2}\}$.
\item Update $\phi_{1:24,1:17}$ using the PDA-MCMC algorithm. We used eight proposals in parallel using parallel MCMC as in \citet{nicholls}. We set
\begin{linenomath}
\begin{equation*}
k_t(M_t)=\prod_{i=1}^{17}N(\ln(w_{t,i})|c(M_t)_i,\sigma_{c,i}^2)
\end{equation*}
\end{linenomath}
for $t=1,\ldots,6$ and
\begin{linenomath}
\begin{equation*}
k_t(M_t)=\prod_{i=1}^{17}N(\ln(w_{t,i})|c(M_t)_i,\sigma_{c,i}^2)N(\ln(z_{t,i})|s(M_t)_i, \sigma_{s,i}^2)
\end{equation*}
\end{linenomath}
for $t=7,\ldots{},24$.
\label{step:2}
\item We updated $\ln{\kappa}$ and $\ln{R_{max,1:17}}$ by proposing several alternatives and moving between them using Calderhead's parallel MCMC algorithm \citep{caulderhead}.
\item We updated $\bm\sigma_c^2$ and $\bm\sigma_s^2$ using Gibbs samplers.
\end{enumerate}
For a description of Cui et al.'s and Calderhead's parallel MCMC see the Supplementary material (S2).

\section{Results}
\label{sec:results}
The MCMC algorithm was run for 20,000 iterations, dropping the initial 10,000 as burn-in. The convergence of the MCMC was checked visually by examining the traceplots of the parameters (see Supplementary material (S4) for traceplots and results of the history matching).

\subsection{Posterior distributions}
\begin{figure}[H]
\includegraphics[scale=0.8]{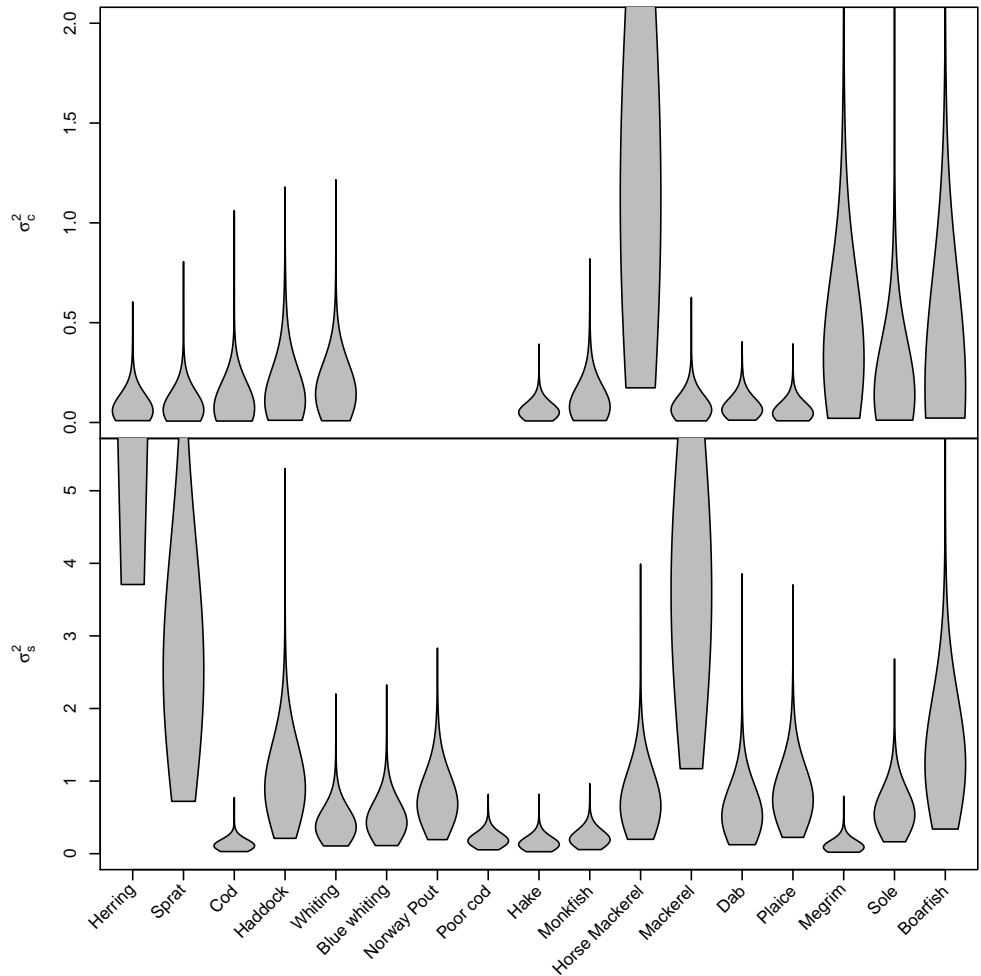}
\caption{Violin plots showing the marginal posterior distribution for the variance parameters. The top plot shows the variance associated with the catch and the bottom shows the variance associated with the survey. Blue whiting's variance term for the catch was large and therefore was omitted from the plot. In the top plot, we fixed $\sigma_c=2$ for Norway Pout and poor cod so they have been omitted from the results.}
\label{fig:var}
\end{figure}

\begin{figure}[H]
\includegraphics[scale=0.8]{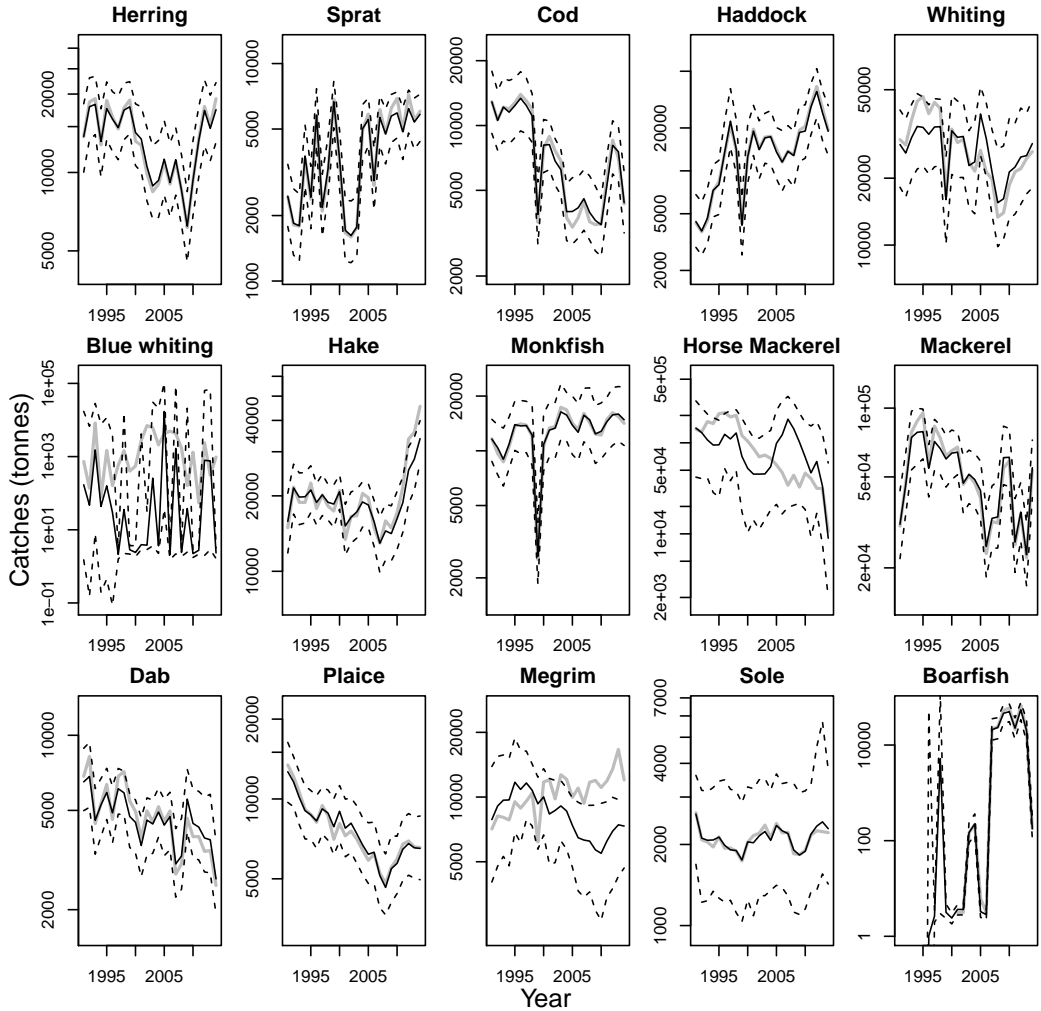}
\caption{The median modelled commercial catches (solid black line), the 10th and 90th percentiles (dotted black lines) and the observed catches (grey line) for 15 of the 17 species. Norway pout and poor cod have been omitted as the model was not fitted to their landings. The downward spike in landings in 1999 for cod, haddock, whiting and monkfish was caused by the French not reporting landing of these stocks in that year in the dataset \citep{ices_landings}.}
\label{fig:catPlot}
\end{figure}

Figure \ref{fig:var} shows the variance parameters for the catches and the survey. The variance parameters describe the estimated distribution of the error around the observed catches as well as the model's inability to predict them. The variance parameters for the catches were much lower than for the survey, particularly for pelagic species, suggesting that the model does a much better job of fitting to commercial catches than the survey data. %This is something that we may expect as the survey is not efficient at catching pelagic fish \citep{walker_selection} and therefore the catch is a lot noisier.
The model does a good job of capturing the catches of most fish with the exceptions of horse mackerel and blue whiting. 
This can also be seen in Figure \ref{fig:catPlot} where we show the median, 10th percentile and 90th percentile of the modelled commercial catches compared to the observed landings (see Supplementary material (Figure S16) for a the same plot for the survey catches). 

\begin{figure}[H]
\begin{center}
\includegraphics[scale=0.6]{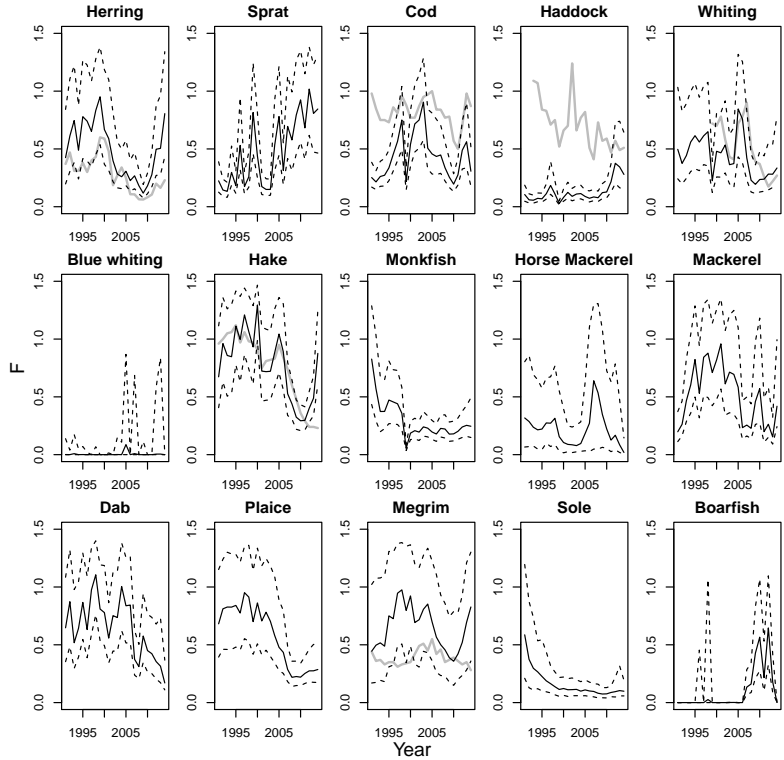}
\end{center}
\caption{The median value of the fishing rates (solid black line), and the 10 and 90 percentiles, (dotted black lines) for 15 of the 17 species. Norway pout and poor cod have been omitted as the model was not fitted to their landings.}
\label{fig:fs}
\end{figure}

Figure \ref{fig:fs} shows the posterior $\phi_{1:17,1:24}$ values for each of the species except Norway pout and poor cod. It also shows the fishing mortality values from the ICES stock-assements, which use SSM, for cod, haddock, whiting, hake, megrim and herring.
The cod, haddock and whiting assessments are for the Celtic Sea \citep{cod_cs18,had_cs18,whi_cs18}, whereas the hake, %assessment,
megrim and herring assessments are for a larger region than our study \citep{hake_cs18,her18,meg18}.
With the exception of haddock, the $\phi_{1:17,1:24}$ values from this study seem to follow, at least qualitatively, that of the assessment fishing mortality.

\begin{figure}[H]
\begin{center}
\includegraphics[scale=0.6]{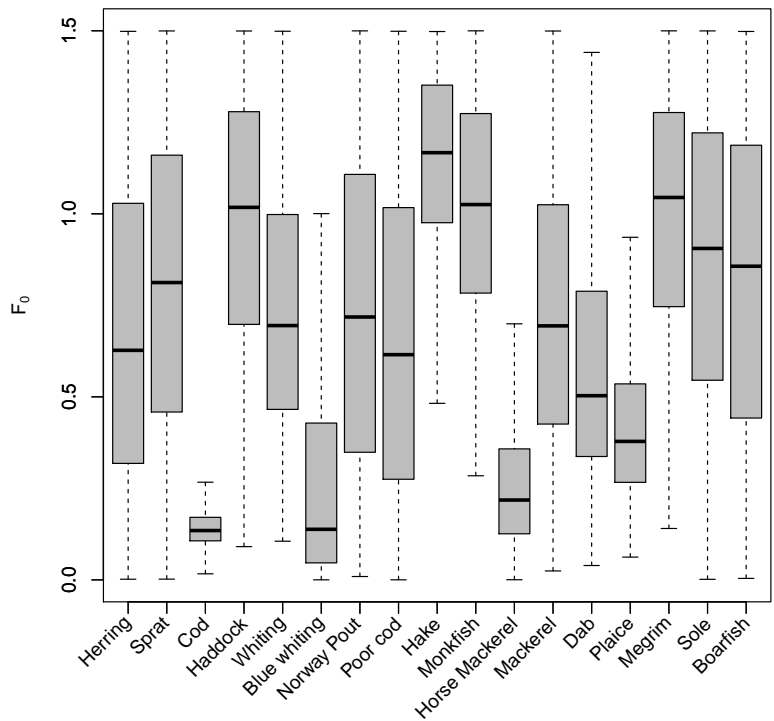}
\end{center}
\caption{The marginal posterior distributions of the fishing rate during the spin-up period, $\phi_0$.}
\label{fig:f0}
\end{figure}

Figure \ref{fig:f0} shows the marginal posterior distribution of the fishing rate during the spin-up period, $\phi_{i,0}$. Many of the posterior distributions are similar to their prior distributions, e.g.\ herring, sprat, however some of the posteriors are quite different from their priors. The fishing rates for cod and horse mackerel are low, which means that when the simulation starts in 1991, cod and horse mackerel will be in a nearly unfished state whereas hake and monkfish, which have quite high fishing rates in the spin-up period, start the simulation in an exploited  state.

\subsection{Spawning stock biomass}
\begin{figure}[H]
\begin{center}
\includegraphics[scale=0.6]{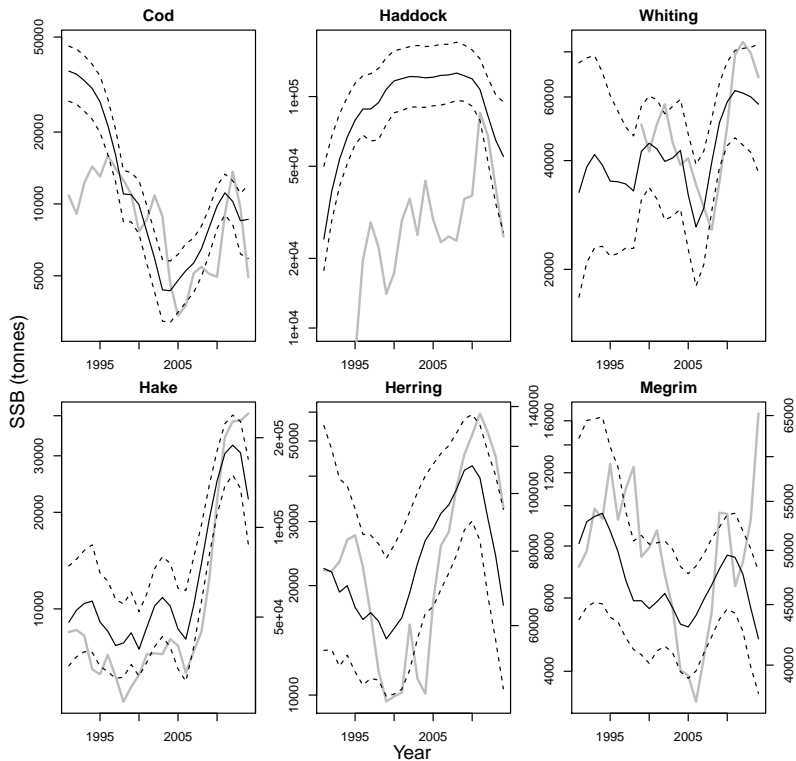}
\end{center}
\caption{The median modelled SSB (solid black line), the 10 and 90 percentiles (dotted black lines) and SSB estimates from single-species ICES assessments for cod, haddock, whiting hake, megrim and herring (grey line). The hake, megrim and herring assessments cover more area than the model does and therefore is plotted on a different scale.}
\label{fig:sa}
\end{figure}

Figure \ref{fig:sa} shows the median, 10th percentile and 90th percentile estimates for cod, haddock, whiting, hake, herring and megrim spawning stock biomass (SSB). It also shows the SSB estimates from ICES stock-assessments. 
The cod assessment and the mizer model agree towards the end of the time period. The whiting single-species and multi-species estimates are similar. Both hake assessments show an increase in SSB at about 2005 which coincides with a reduction in the fishing rate at around the same time, as shown in Figure \ref{fig:fs}; this is also visible in the stock-assessment. In addition the qualitative patterns in herring and megrim seem similar in both the MSM and the SSM. The MSM predicts different SSB for haddock than the SSM.

\section{Discussion}
\label{sec:discussion}
In this study we fitted the multi-species size-based model of \citet{blanchard} with 17 species in the Celtic Sea, a mechanistic model of intermediate complexity, using novel techniques to address the high dimensionality of the problem. We also demonstrated these methods in a simulation study with three species using the model of \citet{LeMaRns}, also a mechanistic MSM (see Simulation study).

\subsection{Mechanistic models}

We found that the model was able to recreate demersal survey catches and commercial catches. The model was not able to recreate the survey data for pelagic fish. This is understandable as the IBTS survey is not so good at sampling pelagic and flatfish and therefore the noise is much greater  \citep{walker_selection}. Our approach gives an idea about the magnitude of the observation uncertainty in the IBTS survey. We could further reduce uncertainty in the model by fitting to additional surveys, for example acoustic surveys.

%% Data

For most of the stocks with full assessments, we get similar SSB and fishing rates, however for haddock both are qualitatively and quantitatively different. In the SSM model the recruitment rates of haddock are unpredictable \citep{had_cs18}, something that is not captured by the MSM model here, which suggests that the SSB in SSMs is recruitment driven. Stochastic recruitment has been included in some MSMs \citep[e.g.][]{blanchard,thorpe17}, but more work is required to explore this.

Although there is such a thing as a true fishing mortality, using it as a `tuning parameter', as done in this study and in many SSMs, destroys its true meaning \citep{rougier2013}. For example, in the model we fitted in this study, only the fishing rates were used to drive the dynamics. Therefore, the fishing rates implicitly have information about all things that drive the dynamics of the species, e.g.\ environment, recruitment or migration. Although many SSMs account for dynamic recruitment \citep[e.g. Stock Synthesis][]{METHOT201386}, their fishing mortality also imply dynamics caused by interactions between different species, which is explicit in MSMs. Therefore taking fishing mortality values from other models and using them as `input variables' \citep[e.g.][]{thorpe15,spence_ns,fishsums}, can lead to systematic biases in the model \citep{brooks} and so should be done with caution, however there are circumstances when it might actually be desirable. For instance it may be necessary to save on computational effort or one may want the fishing rates to represent the fishing mortality generated by stock assessments rather than the actual fishing mortality on the stock as it is possible to calculate this and manage to it \citep[e.g.][]{spence2020fish}.

A common requirement of fisheries models is to assess the current state of a stock. SSMs and statistical MSMs, with many `tuning parameters', are good at doing this when there is a lot of data. However, by fitting mechanistic MSMs directly to data, we free the model from the assessment-induced biases and could therefore contribute to the assessment processes. The natural mortality rates from mechanistic models could be used as `input variables' to SSMs in regions where there is a lack of data (e.g.\ stomach contents data), making statistical MSMs impractical.  For example, results from this Celtic Sea model could be used to generate natural mortality rates that could be used as inputs to SSMs, as currently natural mortality inputs for many of the Celtic Sea assessments come from a theoretical study \citep{Lorenzen_96}. For regions where statistical MSMs already exist, mechanistic MSMs could be used to corroborate or validate them, increasing our confidence in their results, to suggest an alternative or as part of an ensemble model. 

Mechanistic models have been increasingly used as strategic tools when considering how populations, communities and ecosystems respond to management or environmental changes \citep{Pikitch346,Collie_et_al_16}.  They are developed with ecological and biological theory, through `input variables' and processes within the model. Therefore, as this theory develops the mechanistic models become more like reality. As mechanisms and physical laws are time invariant and more robust than statistical correlations, mechanistic MSMs should enable us to make better long-term predictions as interactions between different species and different processes will be more explicit \citep{connor,Cuddington13}. This should lead to improved strategic management, for example in setting long-term targets and reference points, e.g.\ multi-species maximum sustainable yield. Improvements in our understanding of responses to new conditions, such as warming oceans, can readily be included in these models \citep[e.g.][]{Serpetti2017} and the types of actions that can be tested and implemented can be increased, e.g. spatial planning using spatially explicit mechanistic models \citep[e.g. Ecospace ][]{Walters1999}.

Before this study, fitting MSMs to species that did not have full assessments with absolute values of the fishing mortality was not possible without making strong assumptions about their fishing mortality values. This would be particularly the case for species with limited data \citep{QUEROU2013192}. The methods of fitting dynamical parameters introduced and demonstrated here could lead to an increase in the number of mechanistic models for regions where there is not a great amount of information, hence increasing their utility and the strategic management of these areas. This could either be by sharing fishing rates between models, for example a LeMans model for the Celtic Sea could use fishing rates from this study, or directly fitting the dynamical parameters for the mechanistic model. 

In addition, mechanistic models could be used to manage data-limited stocks or in areas of the world where there are many species and building a MSMs is computationally expensive or managing at the level of individual species is impracticable. The methods developed here to find dynamical parameters could be useful when fitting trait-based models, where groups of species with similar traits are grouped together \citep{Barnett_19}.

%%% more than just fishing rates --- any dynamical parameter

%%he advantage of using mechanistic models is that 

\subsection{Developing the role of mechanistic models}
Whilst mechanistic models are potentially powerful tools, their use to date in the advisory process has been limited. Here we suggest some improvements that should make them more useful to fisheries management.

%% Initial conditions 
In this work the state of the system at the beginning of the simulation, $M_0$, was determined by running the model for 300 years with a fixed fishing mortality $\phi_{0,1:17}$, known as the spin-up period \citep{spence_ns}. This led to the model starting in a stationary state, something which may not be true and can have an effect on the results of the model, particularly at the beginning of the simulation. For example, cod was probably not in a stationary state in 1991, as prior to the model large landings were reported in 1988-1990 \citep{cod_cs18}. It is not possible to create the effect of these high landings using the spin-up period, and our fitted model is therefore unable to pick up the dynamics at the beginning of the time series. The fitted model found that the spin-up fishing mortality for cod, $\phi_{0,3}$, was low (Figure \ref{fig:f0}), which lead to over-estimating the SSB (Figure \ref{fig:sa}) and the fishing mortality (Figure \ref{fig:fs}) in the early part of the simulation.

More work is required calibrating the initial state of mechanistic MSMs. In this study we initialised the model by running it model to equilibrium with a fixed fishing mortality. This meant that the system was in equilibrium at the beginning of the simulation which may not be true.
One may run some dynamics, say ten years, before calibration, however it would not have been possible here as we do not know the fishing mortality rates for 1981-1991; alternatively one could run the fishing mortality time series backwards before starting to fit the data, as done in climate modelling \citep{Stouffer2004}. 
A common approach in other fisheries models is to treat the initial states as uncertain, i.e. treating the density for each species and the background for all sizes in mizer as uncertain parameters. We believe this would be the ideal solution, however it would lead to an impractically large number of parameters. A more practical solution may be to use ecological theory from other studies, such as fishing effects on the size-spectrum \citep[e.g.][]{Zhang18}, to parameterise, with only a handful of parameters, the initial state of the model. These parameters would then be calibrated to the data as well.

%% Data

%new bit

%% end

%%% How these mechanistic models can be useful for management

%% assumptions
In this work we used the default fishing selectivity in mizer \citep{mizer}. Other fishing selectivity functions, such as logistic or dome shaped, may lead to different results, however we do not believe that the results would greatly change here.
In the future we would like to include fisheries information, such as effort and catch by fleet or metier, and possibly by size, when fitting these models. In addition information from external studies about the selectivity of different fishing gears could be included, with the selectivity of each gear on each species being the `tuning parameters' \citep[e.g.][]{walker_selection}. One may anticipate that this may follow a dynamic stochastic process, such as an auto-regressive model, as the spatial distribution of species will not change that quickly, thus incorporating more information in the model.

%% validation
With mechanistic MSMs it is not straightforward to perform conventional model validation. In the study here it was not possible to compare the model forecasts with independent out-of-sample data, e.g. the survey and commercial catches in 2015-2019, as the fishing rates, the inputs that are used to drive the dynamics that led to these data, are uncertain. Furthermore, due to the time taken to fit these models it is not practical to perform one-step-ahead analysis \citep{Berg_16} or cross-validation tests. Instead we demonstrated through residual analysis that the conditionally independent assumptions are not violated (see Supplementary material (S4)). There are many other methods that could be used for model validation \citep[e.g.\ posterior predictive checks, see] [for more details]{gelman}; for a recent review of these methods see \citet{Conn18}.

%Common management strategies are to control fishing levels. Current ICES advice is to manage using a constant fishing mortality, the maximum sustainable yield \citep{ices_advice}. 
%As the fishing rates are model-specific, and not the same as the true fishing mortality, it may make more sense to manage an emergent property of the system, such as the biomass \citep{Farcas_Rossberg}.

%The F values that are used to drive the model are model specific. This could lead to problems when running fishing or management scenarios across models or even when implementing them. One possibility may be to manage using the biomass rather than F (Axels paper).
 %The SSB values for cod seem to
%- we may not expect these values to be the same.

%% new bit 

%% single species and statistical MSMs are good for assessment

\subsection{Quantifying uncertainty}
For models to be useful for management it is important that uncertainty is quantified \citep{harwood_stokes}. 
By fitting the model in a Bayesian framework we were able to quantify the uncertainty in the model. This is a difficult problem using conventional MCMC due to the complexity of the model, and the increased dimension of the uncertain parameters caused by fitting fishing mortality. We believe that this is a major reason why this has not previously been done.
SSMs and statistical MSMs take advantage of recent software developments and are fitted using algorithms that exploit gradients, such as Hamiltonian Monte Carlo \citep{neal_hmc} or Reimann Manifold MCMC \citep{girolami}.
However, for more mechanistic-based models, this may be impractical or even impossible. In this paper we have demonstrated a method of exploiting the stucture of the model to use an MCMC algorithm to fit a mechanistic model.

For mechanistic-based models, where the model needs to be run to evaluate the likelihood, it is advantageous to use parallel computing, running several likelihood evaluations at once, to speed up the fitting process. The problem here is that MCMC is a sequential algorithm and therefore difficult to run in parallel \citep{parallel_mcmc}.
In this paper we introduce two novel variations of the delayed-acceptance MCMC algorithm \citep{sherlock}. The MDA-MCMC algorithm is designed to use parallel computing and is motivated by attempting to move many parameters at once, accepting the good moves whilst rejecting the bad ones. We believe that the MDA-MCMC would be most useful when sets of parameters, or transformations of the parameters, affect different parts of the likelihood. This could be explored using variance-based sensitivity analysis \citep{vbsa} prior to running the algorithm. 
% added bit
As the MDA-MCMC algorithm makes moves in smaller dimensions, the proposals can be larger in the parameter space. We recommend the proposals are large so that the resulting in acceptance probabilities, in the first part of the algorithm, are either 0 or 1. This would mean that the accepted points result in large improvements in the full likelihood.

Similarly the PDA-MCMC is motivated by proposing moves in a large number of dynamical parameters but efficiently accepting only the good moves. If one was going to fit the dynamical parameters by hand, one might wish to change the fishing rates one year at a time and to run that model for one year. The PDA-MCMC algorithm does just that but in such a way that the stationary distribution of the Markov chain is the posterior distribution. An alternative would be to change one year at each iteration of the MCMC chain, therefore requiring 24 model runs all of which are required to be done sequentially, whereas using the PDA-MCMC algorithm it only requires five model runs, most of which can be run in parallel. This therefore leads to more efficient use of computational effort when updating dynamical parameters such as annual fishing rates. 
% added bit
The PDA-MCMC algorithm can also be flexible when deciding which of the dynamical parameters are changed. In the study in the manuscript we attempted to change all of the dynamical parameters at once, however in the Simulation study we only changed a handful of dynamical parameters at a time, something that we found led to better mixing.
The PDA-MCMC algorithm is also useful when the state of the model is dependent on the entire past and/or is stochastic. To do this one would require $M_t$ to include the whole of the past. If the model was stochastic, we recommend treating the stochastic elements as additional parameters, as in \citet{Spence2016} allowing better exploration of the dynamical parameter space.
These two algorithms are not specific to MSMs, or even mechanistic models, but are applicable to a wide range of MCMC problems.

%Although an MCMC algorithm will eventually reach stationary distribution no matter where it is started from \citep{gelman}, it may take a long time. If evaluating the likelihood is a slow process, as is often the case for process models, this may not be possible in a realistic time. In this study we used history matching to find initial starting values close to the posterior distribution \citep{veron}. This involves running the model, and building statistical models of the process model, to exclude parameter space that has negligible posterior density. In \citet{spence_ns}, they search around the space keeping the best of their points to further search around. This adds an element of luck, as you need to find a point that is close to the posterior distribution by chance. Therefore it is better to reject implausible parameter space and start the MCMC in what is left.

% -- remove?

%and built statistical models to relate the parameters to the mean catches. We could then exclude regions of space that predicted mean catches that were significantly different from the observed mean catches. `Significantly different' is well defined in History matching () and this is very important there as, the class of problems that history matching is used for often have models that can only be run a few hundred times and thus discounting space means that space is excluded for the rest of the analysis. This is not necessarily in our case as we are only excluding space for the screening process and this space can come back in during the MCMC part.

\subsection{Conclusion}
We have demonstrated a method of fitting mechanistic MSMs directly to data without using SSMs. By using novel techniques we were able to fit a model of intermediate complexity in a high-dimensional parameter space with quantifiable uncertainty.
Furthermore, by fitting mechanistic MSMs directly to data, we free the model from the assessment-induced biases, which may lead to a greater reliability and trust in mechanistic models, increasing their utility in the management process.

Although demonstrated on two multi-species marine models, this methodology is readily generalisable for fitting models of intermediate complexity (with a typical run time of 1 second to a few minutes), when there are a significant number of uncertain dynamic parameters. It is therefore likely to find wide applications throughout science.

\section*{Acknowledgments}
The work was supported by the Natural Environment Research Council and Department for Environment, Food and Rural Affairs [grant number NE/L003279/1, Marine Ecosystems Research Programme]. The authors would like to thank Christopher Lynam, Rui Vieira, Johnathan Ball, Paul Dolder and Christopher Griffiths for comments on an earlier version of the paper. We would also like to thank Paul Hart and two anonymous reviewers for their invaluable contributions. In particular we would like to thank one of the reviewers who meticulously went through the manuscript and supplementary material.

\section*{Authors' contribution}
MAS conceived the ideas and designed the methodology; MAS, RTB, and PGB fit the model to the data; RS, FS and JLB developed the model used in the case study; MAS led the writing of the manuscript. All authors contributed critically to the drafts and gave final approval for publication

\section*{Data availability statement}
Data sharing is not applicable to this article as no new data were created; rather, data were acquired from existing published sources (all sources are cited in the text), or are described, figured and tabulated within the manuscript or supplementary information of this article.

%\bibliographystyle{plainnat}
%\bibliographystyle{mee}
%\bibliography{bibfile}

\end{document}